\documentstyle[12pt,fleqn,epsf]{article}
\newtheorem{theorem}{Theorem}
\newtheorem{proposition}{Proposition}
\newtheorem{lemma}{Lemma}
\newtheorem{definition}{Definition}

\renewcommand{\arraystretch}{1.3}

\def\nsection#1{\setcounter{equation}{0}\section{#1}}

\newcommand{\Integer}{\:\mbox{\sf Z} \hspace{-0.82em} \mbox{\sf Z}\,}

\def\Multb#1#2{\left[#1 \atop #2 \right]}
\def\Mult#1#2{\biggl[{#1 \atop #2}\biggr]}
\def\Mults#1#2{\left[{\textstyle {#1 \atop #2} } \right]}
\def\e{\mbox{e}}
\def\es{\mbox{\scriptsize e}}
\def\case#1#2{{\textstyle{#1\over #2}}}
\def\eps{\varepsilon}
\def\la{\lambda}
\def\mod#1{\; (\bmod \: #1)}

\begin{document}

\title{Fermionic solution of the Andrews-Baxter-Forrester model II:
proof of Melzer's polynomial identities}

\author{S.~Ole Warnaar\thanks{
e-mail: {\tt warnaar@maths.mu.oz.au}}
\\
Mathematics Department\\
University of Melbourne\\
Parkville, Victoria 3052\\
Australia}

\date{August, 1995 \\ \hspace{1mm}
\\
Preprint No. 17-95}
\maketitle

\begin{abstract}
We compute the one-dimensional configuration sums
of the ABF model using the fermionic technique
introduced in part I of this paper.
Combined with the results of Andrews,
Baxter and Forrester, we find proof of polynomial
identities for finitizations of the
Virasoro characters $\chi_{b,a}^{(r-1,r)}(q)$ as
conjectured by Melzer.
In the thermodynamic limit these identities
reproduce Rogers--Ramanujan type identities
for the unitary minimal Virasoro characters,
conjectured by the Stony Brook group.
We also present a list of additional Virasoro character identities
which follow from our proof of Melzer's identities and application
of Bailey's lemma.

\end{abstract}

\newpage

\nsection{Introduction}
Probably among the most celebrated results in
mathematics are the identities of Rogers and
Ramanujan~\cite{Rogers94,Rogers19,Ramanujan}
\begin{equation}
\sum_{m = 0}^{\infty} \frac{q^{m(m+a)}}{(q)_{m}}
=\frac{1}{(q)_{\infty}}\sum_{j=-\infty}^{\infty} (-1)^j
q^{j(5j +1+2a)/2} \qquad \qquad a=0,1,
\label{RR}
\end{equation}
where
$(q)_m = \prod_{k=1}^m (1-q^k)$, $m>0$ and
$(q)_0=1$.
In the context of modern physics, one recognizes the right-hand side
of these identities to be the Rocha-Caridi expression for the Virasoro
characters $\chi^{(2,5)}_{1,2-a}(q)$ of minimal conformal field theory
$M(2,5)$~\cite{Rocha}.
As such, the Rogers--Ramanujan identities can be seen
as character identities of some Virasoro algebra. A natural question
is whether the other Virasoro characters also admit
identities of the Rogers--Ramanujan type.
For the important
class of {\em unitary} minimal models $M(r-1,r)$,
this was answered affirmative in a remarkable paper by the
Stony Brook group~\cite{KKMMa}.\footnote{By now character identities
of Rogers--Ramanujan type for all minimal Virasoro characters
$\chi^{(p,p')}(q)$ have been found~\cite{KKMMa,FQ,BM}.}
However, the results of ref.~\cite{KKMMa}
were all based on extensive numerical studies, and actual
proofs remained elusive.

Among the many methods of proof of the original Rogers--Ramanujan
identities an elegant approach is that of first proving
the polynomial identities \cite{Schur,Andrews}
\begin{equation}
\sum_{m=0}^{\infty} q^{m(m+a)}
\Multb{L-m-a}{m} =
\sum_{j=-\infty}^{\infty} (-1)^j
q^{j(5j+1+2a)/2}
\Multb{L}{\lfloor \frac{1}{2}(L -5j-a)\rfloor},
\label{finiteRR}
\end{equation}
for all $L\geq a$.
Here $\lfloor x \rfloor$ denotes the integer part of $x$
and
$\Mults{N}{m}$ is a Gaussian polynomial
defined as
\begin{equation}
\renewcommand{\arraystretch}{1.5}
\Mult{N}{m} = \left\{
\begin{array}{ll}
\displaystyle \frac{(q)_N}{(q)_{m}(q)_{N-m}} \qquad & 0\leq m \leq N \\
0& \mbox{otherwise.}
\end{array}
\right.
\label{qpoly}
\end{equation}
Clearly, in the limit of $L\to\infty$ we recover the Rogers--Ramanujan
identity (\ref{RR}).
To proof the {\em finitized} Rogers--Ramanujan identities (\ref{finiteRR})
it suffices to check that both left- and right-hand side satisfy
the elementary recurrences $\mbox{$f_L=f_{L-1}+q^{L-1} f_{L-2}$}$
as well as the same initial conditions for $L=a,a+1$.

In an attempt to find proofs of the
identities for the characters $\chi^{(r-1,r)}_{b,a}(q)$ (see next
section for their actual form), Melzer followed Schur's approach
and conjectured
finitizations similar to those in (\ref{finiteRR}).
However, Melzer's polynomial identities were sufficiently complicated
not to lead to a straightforward proof using recurrences.
It was only after Melzer proved the cases
$r=3$ (Ising) and $r=4$ (tricritical Ising)~\cite{Melzer}
that Berkovich succeeded  in proving recurrences for the polynomial
identities for all $\chi^{(r,r-1)}_{b,1}(q)$~\cite{Berkovich}.

In this paper we present a combinatorial proof for
Melzer's identities, based on yet another observation made by
Melzer. Again the motivation for this has been the original
Rogers-Ramanujan identities (\ref{RR}), whose finitization
(\ref{finiteRR}) can be viewed as evaluations of the sum
\begin{equation}
\renewcommand{\arraystretch}{1}
\sum_{
\begin{array}{c} \scriptstyle
\sigma_1,\ldots,\sigma_{L-1}=0,1\\
\scriptstyle
\sigma_j\sigma_{j+1}=0
\end{array}}
q^{ \;\; \displaystyle \sum_{k=1}^{L-1} k\sigma_k} \qquad
\qquad  \sigma_0=a, \; \sigma_L=0,
\label{HHM}
\end{equation}
in two intrinsically different ways.
Similar to this, Melzer has argued that the polynomial identities for
the finitized $\chi^{(r-1,r)}_{b,a}(q)$ characters arise from
computing the sums
\begin{equation}
\renewcommand{\arraystretch}{1}
X_L(a,b) = \!\!\!\sum_{
\begin{array}{c} \scriptstyle
\sigma_1,\ldots,\sigma_{L-1}=0 \\
\scriptstyle
|\sigma_{j+1}\!-\!\sigma_j|=1
\end{array}}^{r-2} \!\!
q^{\;\;\displaystyle \sum_{k=1}^L
k|\sigma_{k+1}-\sigma_{k-1}|/4}
\quad  \sigma_0=a-1, \; \sigma_L=b-1, \; \sigma_{L+1}=b,
\label{confsums}
\end{equation}
for all $a=1,\ldots,r-1$ and $b=1,\ldots,r-2$.

We will take this observation as the starting point for proving
the polynomial and Rogers--Ramanujan identities for the
(finitized) characters $\chi^{(r-1,r)}_{b,a}(q)$.
That is, we give two different methods to compute (\ref{confsums}),
one leading to a so called {\em fermionic} expression similar to
the left-hand side of (\ref{finiteRR}) and one method leading to a
so-called {\em bosonic} expression similar to the right-hand side
of (\ref{finiteRR}).
In fact, it should be noted that $X_L(a,b)$ defined above is
exactly the {\em one-dimensional configuration sum} $X_L(a,b,c)$, with
$c=b+1$, as defined by Andrews, Baxter and Forrester in their computation
of the order parameters of the $(r-1)$-state ABF model in regime III
\cite{ABF}.
Hence computing the sum (\ref{confsums}) amounts to computing
the order parameters of the ABF model. The fact that (finitized)
Rogers--Ramanujan identities arise from calculating order parameters
of solvable lattice models is in fact not new, and indeed the sum
(\ref{HHM}) is exactly the one encountered by Baxter in his
solution of the hard hexagon model in regime~I~\cite{Baxter81}.

The remainder of this paper is organized as follows.
In the next section we describe Melzer's polynomial identities,
their limiting Rogers--Ramanujan type form and some other
Virasoro character identities that follow from the proof of
Melzer's identities and the application of the Andrews--Bailey
construction~\cite{Bailey47,Bailey49,Andrews84}.
Then, in section~3, we compute the configuration sums $(\ref{confsums})$
using the technique developed in part I of this paper \cite{W}.
This amounts to reinterpreting the sum (\ref{confsums}) as the
grand canonical partition function of a one-dimensional gas of
charged particles obeying certain Fermi-type exclusion rules.
In section~4 we describe the original approach of ABF for computing
(\ref{confsums}) using recurrence relations.
Together with the result of section~3 this proves Melzer's
polynomial identities.
We finally end with a discussion of our result and an outlook to
related problems and generalizations.

To end this introduction we make some further remarks
on the problem described in this paper.
First, as mentioned before, an altogether different kind
of proof of Melzer's identities has recently been given for
the case of $\chi^{(r-1,r)}_{b,1}(q)$ by Berkovich \cite{Berkovich}.
This method of proof, which in fact is applicable to all
unitary minimal characters \cite{BM}, is based on recursive
instead of combinatorial arguments.\footnote{Berkovich
has subsequently  proven Melzer's identities for all characters,
but his results remain unpublished \cite{Berkovich2}.}

Second, in their solution of the ABF model, Andrews, Baxter
and Forrester also considered the configuration sums $X_L(a,b,c)$,
with $c=b-1$. Hence to completely compute all configuration sums
of the ABF model, more general sums than those defined in
(\ref{confsums}) have to be considered. However, from simple
symmetry arguments \cite{ABF,Melzer} (see also section 3) one can
easily deduce that computing (\ref{confsums}) suffices to obtain
expressions for all $X_L(a,b,c)$.

Finally we remark that Melzer~\cite{Melzer} and
Kedem {\em et al.}~\cite{KKMMa} conjecture
(in the general case) four fermionic expressions for each
(finitized) character.
In this paper we give detailed proof of only two of the four.
For the remaining two representations we did not succeed
in finding a derivation in terms of a Fermi lattice-gas.

\nsection{Melzer's polynomial identities and related
Rogers--Ra\-manujan identities}
In this section we give a summary of identities proven by the
calculations carried out in the sections~\ref{sec3} and~\ref{sec4}.
First we describe the polynomial identities conjectured by
Melzer~\cite{Melzer},
and their limiting Rogers--Ramanujan type form as discovered by the
Stony Brook group~\cite{KKMMa}. Then we list two classes of character
identities for non-unitary minimal models which, as recently pointed
out by Foda and Quano \cite{FQ}, arise from Melzer's identities
and the Andrews--Bailey construction~\cite{Bailey47,Bailey49,Andrews84}.

\subsection{Identities for the (finitized) Virasoro characters
$\chi^{(r-1,r)}_{b,a}(q)$}
Before we state the polynomial identities as conjectured by Melzer,
we need some notation.
We denote the incidence matrix of the A$_{r-3}$ Dynkin diagram by
$\cal I$, with ${\cal I}_{j,k}=\delta_{j,k-1}+\delta_{j,k+1}$,
$j,k=1,\ldots,r-3$. The Cartan matrix of A$_{r-3}$ is denoted
as $C$, and is related to $\cal I$ by $C_{j,k}=2\delta_{j,k}
-{\cal I}_{j,k}$. We also define the $(r-3)$-dimensional
(column) vectors $\vec{m}$ and $\vec{\e}_j$, $j=1,\ldots,r-3$,
by $(\vec{m})_j=m_j$ and $(\vec{\e}_j)_k=\delta_{j,k}$, and set
$m_0=m_{r-2}=0$, $\vec{\e}_0=\vec{\e}_{r-2}= \vec{0}$.
With this notation, using the Gaussian
polynomials as defined in (\ref{qpoly}), Melzer's conjectures
can be stated as the following identities for
$a=1,\ldots,r-1$, $b=1,\ldots,r-2$ and
$L-|a-b|\in 2\Integer_{\geq 0}$:\footnote{Throughout this paper
we use the notation $x \equiv y$ to mean $x \equiv y \mod{2}$
Also, the sums $\sum_{\vec{x} \equiv \vec{y}}$ and $\sum_{\vec{x}}$
are shorthand notations for $\prod_j \sum_{x_j \geq 0; \;
x_j \equiv y_j}$ and $\prod_j \sum_{x_j \geq 0}$,
respectively.}
\begin{eqnarray}
\lefteqn{
f_{a,b}
\sum_{\vec{m} \equiv \vec{Q}_{a,b}}
q^{\, \case{1}{4} \,  \vec{m}^T C \, \vec{m}
-\case{1}{2} m_{r-a-1} }
\prod_{j=1}^{r-3}
\Mult{\frac{1}{2}({\cal I} \, \vec{m} + L \, \vec{\e}_1 +
\vec{\e}_{r-a-1}
+\vec{\e}_{r-b-1})_j}{m_j} } \nonumber \\
& & =
\sum_{j=-\infty}^{\infty} \left\{
q^{j\big(r(r-1)j+rb-(r-1)a\big)}
\Mult{L}{\scriptstyle \frac{1}{2}(L+a-b)-rj}
-q^{\big((r-1)j+b\big)\big(rj+a\big)}
\Mult{L}{\scriptstyle \frac{1}{2}(L-a-b)-rj} \right\},
\label{Mid}
\end{eqnarray}
with $f_{a,b}= q^{-(a-b)(a-b-1)/4}$ and
\begin{equation}
\vec{Q}_{a,b} = \vec{Q}_{a,b}^{(r-3)} =
(\vec{\e}_{r-a-2}+\vec{\e}_{r-a-4}+\ldots)
+(\vec{\e}_{r-b-2}+\vec{\e}_{r-b-4}+\ldots ).
\label{Qrest}
\end{equation}
We note that in our derivation of the left-hand side of
(\ref{Mid}) in section~3, this restriction naturally arises
in the following form, $\mod{2}$-equivalent to (\ref{Qrest}):
\begin{equation}
(\vec{Q}_{a,b})_j = \min(a-1,r-j-2)+\min(b-1,r-j-2).
\end{equation}

In ref.~\cite{Melzer}, yet another expression for the left-hand side of
(\ref{Mid}) was conjectured as
\begin{equation}
f_{a,b} \sum_{\vec{m} \equiv \vec{R}_{a,b}}
q^{\, \case{1}{4} \,  \vec{m}^T C \, \vec{m}
-\frac{1}{2} m_{a-1} }
\prod_{j=1}^{r-3}
\Mult{\frac{1}{2}({\cal I} \, \vec{m} + L \, \vec{\e}_1 +
\vec{\e}_{a-1} +\vec{\e}_{r-b-1})_j}{m_j}
\label{Mid2}
\end{equation}
where
\begin{equation}
\vec{R}_{a,b} = (r-a-1)\vec{\rho}+(\vec{\e}_{a}+\vec{\e}_{a+2}+\ldots)
+(\vec{\e}_{r-b-2}+\vec{\e}_{r-b-4}+\ldots ),
\end{equation}
with $\vec{\rho}=\sum_{j=1}^{r-3} \vec{\e}_j$.
Clearly, for $a=1$ and for $a=r-1$ the fermionic expressions in
(\ref{Mid}) and (\ref{Mid2}) coincide.

As mentioned in the introduction,
we have no explanation of this alternative fermionic
form in terms of a Fermi-gas, and (\ref{Mid2}) is listed
only for completeness.

Taking the finitization parameter $L$ to infinity, (\ref{Mid})
leads to Rogers--Ramanujan type identities for unitary minimal
Virasoro characters. Hereto we recall the well-known
Rocha-Caridi expression for all (normalized)
characters $\chi^{(p,p')}_{r,s}(q)$ of minimal CFT $M(p,p')$,
\begin{equation}
\chi^{(p,p')}_{r,s}(q)=\frac{1}{(q)_{\infty}}
\sum_{j=-\infty}^{\infty} \left\{
q^{j(pp'j+p'r-ps)}-q^{(jp+r)(jp'+s)} \right\},
\label{RC}
\end{equation}
for $r=1,\ldots,p-1$, $s=1,\ldots,p'-1$, with $p$ and $p'$ coprime.
We thus find that the right-hand side of (\ref{Mid}) gives the
{\em bosonic} Rocha-Caridi expression for $\chi^{(r-1,r)}_{b,a}(q)$,
whereas the left-hand side leads to a {\em fermionic} counterpart,
\begin{equation}
\chi^{(r-1,r)}_{b,a}(q) = f_{a,b}
\sum_{\vec{m} \equiv \vec{Q}_{a,b} }
\frac{q^{\, \case{1}{4} \,  \vec{m}^T C \, \vec{m}
-\case{1}{2} m_{r-a-1} } }{(q)_{m_1}}
\prod_{j=2}^{r-3}
\Mult{\frac{1}{2}({\cal I} \, \vec{m} +
\vec{\e}_{r-a-1} +\vec{\e}_{r-b-1})_j}{m_j} .
\label{SB}
\end{equation}
This result is one of the many celebrated conjectures
for fermionic character representations
made by the Stony Brook group, see e.g., refs.~\cite{KKMMa,KKMMb,DKKMM}.

An obvious symmetry of (\ref{RC}) is $\chi^{(p,p')}_{r,s}(q)=
\chi^{(p,p')}_{p-r,p'-s}(q)$. Making the transformation
$a\to r-a$ and $b\to r-b-1$ in the fermionic expression
(\ref{SB}) this symmetry is not at all manifest, except for $b=1$ and
$a=1,r-2$.
Hence we have two different fermionic representations
for each character of the unitary minimal series.

To end our discussion on Melzer's polynomial identities, we remark
that in ref.~\cite{Melzer} identities were also given for
finitizations of the characters $\chi^{(r-1,r)}_{b,a}(q)$,
with finitization parameter $L$ such that $L+a-b \not\equiv 0$.
Since these can simply be obtained from (\ref{Mid}) and (\ref{Mid2})
by the above-mentioned symmetry transformation, they are not listed
here as separate identities.

\subsection{Rogers--Ramanujan identities for
$\chi_{a,(k+1)b}^{(r,(k+1)r-1)}(q)$ and
$\chi_{b,(k+1)a}^{(r-1,(k+1)r-k)}(q)$}
It was recently pointed out by Foda and Quano~\cite{FQ},
that many new Virasoro
character identities can be obtained by applying some powerful lemmas,
proven by Bailey and Andrews, to Melzer's polynomial identities.
The main idea of these lemmas is to proof the more complicated
Rogers--Ramanujan type identities by showing that they are
a consequence of easier to proof identities.
Here we will not state the relevant lemmas but refer
the interested reader
to the work of Foda and Quano~\cite{FQ} and to the original work
of Bailey~\cite{Bailey47,Bailey49} and Andrews~\cite{Andrews84}.

In both series of Virasoro character identities given below,
we encounter the $k$ by $k$ matrix $B$ with entries
$B_{j,\ell}=\mbox{min}(j,\ell)$.
We note that this matrix is the inverse of the Cartan-type matrix
of the tadpole graph with $k$ nodes;
$(B^{-1})_{j,\ell} = 2\delta_{j,\ell}-{\cal I}_{j,\ell}^{(k)}$,
with incidence matrix of the tadpole graph given by
${\cal I}_{j,\ell}^{(k)}=\delta_{j,\ell-1}
+\delta_{j,\ell+1}+\delta_{j,\ell}\,\delta_{j,k}$, $j,\ell=1,\ldots,k$.
We will also use the $k$-dimensional vectors $\vec{n}$ and
$\vec{\eps}_k$, whose $j$-th entries read $n_j$ and $\delta_{k,j}$,
respectively.

\subsubsection{$\chi_{a,(k+1)b}^{(r,(k+1)r-1)}(q)$}
Substituting the {\em Bailey pair} read off from (\ref{Mid}) into the
{\em Bailey chain} of length $k$, we obtain
\begin{eqnarray}
\chi_{a,(k+1)b}^{(r,(k+1)r-1)}(q)
&\stackrel{\vphantom{\case{1}{2}} (a\equiv b)}{=}&
f_{a,b} \; q^{-k(a-b)^2/4}
\sum_{\vec{n}}
\sum_{\vec{m} \equiv \vec{Q}_{a,b}}
\frac{q^{\,\vec{n}^T B \: \vec{n}}}
{(q)_{n_1} \ldots (q)_{n_{k-1}} (q)_{2n_k}}  \nonumber \\
& & \times \;
q^{\, \case{1}{4} \,  \vec{m}^T C \, \vec{m}
-\case{1}{2} m_{r-a-1} }
\prod_{j=1}^{r-3}
\Mults{\case{1}{2}({\cal I} \, \vec{m} + 2n_k \, \vec{\es}_1 +
\vec{\es}_{r-a-1}
+\vec{\es}_{r-b-1})_j}{m_j}
\nonumber \\
&\stackrel{\vphantom{\case{1}{2}} (a\not\equiv b)}{=}&
f_{a,b} \; q^{-k\bigl((a-b)^2-1\bigr)/4}
\sum_{\vec{n}}
\sum_{\vec{m} \equiv \vec{Q}_{a,b} }
\frac{q^{\,\vec{n}^T B \: (\vec{n} +\vec{\eps}_k)}}
{(q)_{n_1} \ldots (q)_{n_{k-1}} (q)_{2n_k+1}}
\label{AB} \\
& & \times \;
q^{\, \case{1}{4} \,  \vec{m}^T C \, \vec{m}
-\case{1}{2} m_{r-a-1} }
\prod_{j=1}^{r-3}
\Mults{\case{1}{2}({\cal I} \, \vec{m} + (2n_k+1) \, \vec{\es}_1 +
\vec{\es}_{r-a-1}
+\vec{\es}_{r-b-1})_j}{m_j}
\nonumber
\end{eqnarray}
valid for all $k\geq 1$, $a=1,\ldots,r-1$, $b=1,\ldots,r-2$.

The proof of this result for $a=1$ was first noted by Foda and
Quano \cite{FQ}, using the proof of Melzer's identities for $a=1$ as
established by Berkovich~\cite{Berkovich}. The fermionic expression
in (\ref{AB}) can also be found in ref.~\cite{BM}.

\subsubsection{$\chi_{b,(k+1)a}^{(r-1,(k+1)r-k)}(q)$}
Substitute the {\em dual} Bailey pair
obtained from (\ref{Mid}) into the
Bailey chain of length $k+1$.  Then make  the change
of variables $m_j\to m_{j+1}$, followed by $2 n_{k+1}+|a-b|\to m_1$,
$n_k \to n_k +\frac{1}{2}(m_1-|a-b|)$ and $r\to r-1$.
Finally, interchanging $a$ and $b$ then using
\begin{equation}
\left( \vec{Q}_{a,b}^{(r-3)} \right)_j \equiv \left\{
\begin{array}{ll}
\left( \vec{Q}_{b,a}^{(r-4)} \right)_{j-1}
\quad & j=2,\ldots,r-3 \\
a-b & j=1,
\end{array} \right.
\end{equation}
true for $a=1,\ldots,r-3$, $b=1,\ldots,r-2$,
yields
\begin{eqnarray}
\chi_{b,(k+1)a}^{(r-1,(k+1)r-k)}(q) &=&
f_{a,b} \; q^{-k(a-b)^2/4}
\sum_{\vec{n} }
\sum_{\vec{m} \equiv \vec{Q}_{a,b} }
\frac{q^{ (\vec{n} + \case{1}{2} m_1 \vec{\eps}_k)^T B \:
(\vec{n} + \case{1}{2} m_1 \vec{\eps}_k)}}
{(q)_{n_1} \ldots (q)_{n_k}}
\nonumber \\
& & \times \;
\frac{q^{\, \case{1}{4} \,  \vec{m}^T C \, \vec{m}
-\case{1}{2} m_{r-a-1} }}{(q)_{m_1}}
\prod_{j=2}^{r-3}
\Mults{\case{1}{2}({\cal I} \, \vec{m} +
\vec{\es}_{r-a-1} +\vec{\es}_{r-b-1})_j}{m_j},
\label{dualB}
\end{eqnarray}
valid for all $k\geq 0$, $a=1,\ldots,r-3$, $b=1,\ldots,r-2$.
Note that for $k=0$, corresponding to a Bailey chain
of length 1, we actually recover a subset of the character
identities (\ref{SB}) for $M(r-1,r)$.

For $a=b=1$, (\ref{dualB}) was conjectured in ref.~\cite{KKMMa}.
The proof for $a=1$ can again be found in ref.~\cite{FQ},
though the actual form of the fermionic side therein
rather differs due to the sequence of the transformations
carried out above. The fermionic form (\ref{dualB})
can also be found in ref.~\cite{BM}.

\nsection{Fermionic solution of the ABF model}\label{sec3}
We now come to the main part of this paper, the
evaluation of the one-dimensional configuration
sums (\ref{confsums}) of the ABF model. This  yields,
up to the prefactor $f_{a,b}$,
the left-hand side of the identity (\ref{Mid}).
To establish this, we first reformulate the sum (\ref{confsums})
as the generating function
of certain restricted lattice paths. We then compute
this generating function by identifying each path
as a configuration of charged fermions on a one-dimensional
lattice. This identification allows us to view $X_L(a,b)$
as the grand-canonical partition function
of a one-dimensional Fermi-gas.
Because of the one-dimensional nature of this gas, its
partition function can readily be computed.

\subsection{Restricted lattice paths}
To reformulate the sum (\ref{confsums}) in terms of lattice paths,
we first give some basic definitions.
\begin{definition}
An ordered sequence of spins
$\{\sigma_0,\sigma_1,\ldots,\sigma_{L+1}\}$ is called
admissible if
\begin{itemize}
\item $\sigma_j\in \{0,1,\ldots,r-2\}$ for $j=0,\ldots,L+1$,
\item $\mbox{$|\sigma_{j+1}-\sigma_j|=1$}$ for $j=0,\ldots,L$, and
\item $\sigma_0=a-1$, $\sigma_L=b-1$ and $\sigma_{L+1}=b$.
\end{itemize}
\end{definition}

\begin{figure}[t]
\epsfxsize = 10cm
\centerline{\epsffile{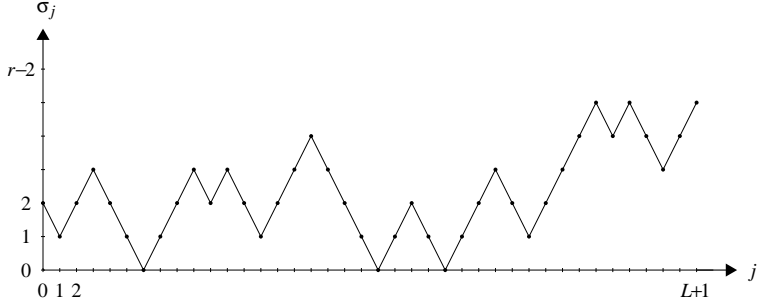}}
\caption{An example of a restricted lattice
path in rlp$(0,r)$.}
\label{fig1}
\end{figure}

\begin{definition}
Let $\{\sigma_0,\sigma_1,\ldots,\sigma_{L+1}\}$ be an admissible
sequence of spins. Plot all pairs $(j,\sigma_j)$
in the $(x,y)$-plane and interpolate
between each pair of neighbouring points by a straight line
segment. The resulting graph is called  a restricted lattice path.
\end{definition}
An example of a restricted lattice path
for $a=3$ and $b=5$ is shown in Figure~\ref{fig1}.

To write the one-dimensional configuration sum
as a sum over restricted lattice paths, first notice that
the restrictions on the $\sigma$'s in
(\ref{confsums}) precisely correspond to
those defining an admissible sequence of spins.
Consequently, each restricted lattice path
corresponds to one of the terms in the sum (\ref{confsums})
and, conversely, each term in the sum corresponds to
a restricted lattice path.
Given an admissible sequence, its total weight
is decomposed as follows. If $\sigma_{j-1}<\sigma_j<\sigma_{j+1}$ or
$\sigma_{j-1}>\sigma_j>\sigma_{j+1}$ this contributes a factor
$q^{j/2}$ and if
$\sigma_{j-1}<\sigma_j>\sigma_{j+1}$ or
$\sigma_{j-1}>\sigma_j<\sigma_{j+1}$ this contributes a factor 1.
In terms of the restricted lattice paths this simply means that
for each integer point $j$ along the $x$-axis we get a factor
1 if $(j,\sigma_j)$ is an extremum and a factor $q^{j/2}$
otherwise. Here the terminals of a path are to be viewed
as extrema.
Writing this in the language of statistical mechanics we get,
setting $q=\exp(-\beta)$,
\begin{equation}
X_L(a,b) = \sum_{\mbox{\scriptsize restricted lattice paths}}
\e^{\; \displaystyle -\beta \sum_{j=1}^L E(j)},
\label{Xpaths}
\end{equation}
with energy function $E$ given by
\begin{equation}
E(j) = \left\{
\begin{array}{ll}
0 & \mbox{if the path has an extremum at ($x$-position) $j$} \\
\frac{1}{2} j
\qquad & \mbox{otherwise.}
\end{array}\right.
\label{energy}
\end{equation}

Each of the lattice paths in the sum (\ref{Xpaths})
starts in $(0,a-1)$, ends in $(L,b-1),\; (L+1,b)$ and
is restricted to the strip $0\leq y \leq r-2$.
We now define rlp$(\mu,r)$ as the set of all restricted
lattice paths with minimal $y$ value equal to $\mu$ and
maximal $y$ value less or equal to $r-2$.
Hence we can write
\begin{equation}
X_L(a,b)=\sum_{\mu=0}^{\min(a,b)-1} \Xi_L(a,b;\mu,r),
\end{equation}
with
\begin{equation}
\Xi_L(a,b;\mu,r) = \sum_{\mbox{\scriptsize rlp}(\mu,r)}
\e^{\; \displaystyle -\beta \sum_{j=1}^L E(j)}.
\label{XiLabmu}
\end{equation}
Noting the obvious relation
$\Xi_L(a,b;\mu,r) = \Xi_L(a-\mu,b-\mu;0,r-\mu)$
gives
\begin{equation}
X_L(a,b)=\sum_{\mu=0}^{\min(a,b)-1} \Xi_L(a-\mu,b-\mu;0,r-\mu),
\label{summu}
\end{equation}
and we conclude that to compute $X_L(a,b)$ it suffices to compute
sum (\ref{XiLabmu}) for $\mu=0$, and arbitrary $a$, $b$ and $r$.

So far we only have reformulated the problem of
computing $X_L(a,b)$, and it is by no means clear that
$\Xi_L(a,b):=\Xi_L(a,b;0,r)$ is any simpler to evaluate than
(\ref{confsums}). To make some real progress,
we will show in the next section that $\Xi_L(a,b)$ can be
viewed as the grand canonical partition function
of a one-dimensional gas of charged fermions.
In other words, each path in rlp$(0,r)$
can be viewed as a configuration of an appropriately
defined Fermi-gas.
Now decomposing the sum over all Fermi-gas configurations
into a sum over configuration with fixed particle
content (FC) and a sum over the particle content (C),
we get
\begin{equation}
\Xi_L(a,b) = \sum_{\mbox{\scriptsize C}}
Z(\mbox{C};a,b),
\label{ZCab}
\end{equation}
with $Z(\mbox{C};a,b)$
the partition function of the 1-dimensional
Fermi-gas,
\begin{equation}
Z(\mbox{C};a,b) = \sum_{\mbox{\scriptsize FC}}
\e^{\; \displaystyle -\beta \sum_{j=1}^L E(j)}.
\end{equation}

\subsection{A one-dimensional Fermi-gas}
To interpret each restricted lattice path in rlp$(0,r)$
as a configuration of particles, we need some more terminology.
In fact, since some of the concepts introduced below are
somewhat awkward to describe, but easily explained pictorially,
we state some definitions purely graphically.

In the previous section restricted lattice path were
introduced as path from $(0,a-1)$ to $(L,b-1)$, $(L+1,b)$,
restricted to the strip $0\leq y \leq r-2$, such that
$y_{j+1}-y_j=\pm 1$ for all consecutive points
$(j,y_j)$ and $(j+1,y_{j+1})$ on the path.
We somewhat relax these conditions by defining a
{\em lattice path} as
\begin{definition}
A lattice path is a restricted lattice path  with
arbitrary (integer) begin- and endpoint.
\end{definition}
In particular, if a lattice path ends in $(j,y_j)$, the
$y$-coordinate of the second-last point can either be
$y_j-1$ or $y_j+1$.

\begin{figure}[t]
\epsfysize = 3.5cm
\centerline{\epsffile{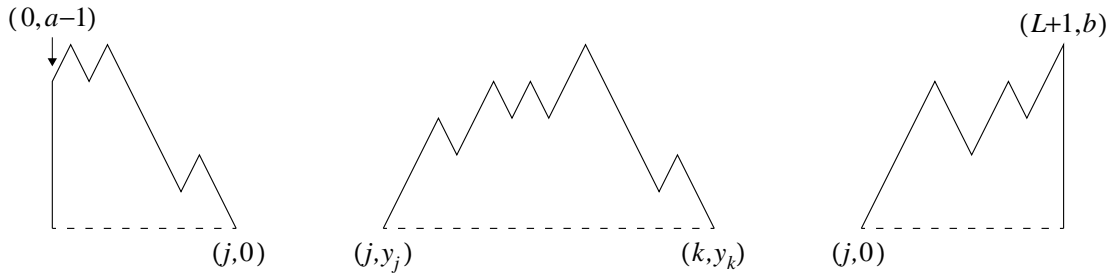}}
\caption{Typical examples of a left-boundary, bulk, and
right-boundary complex.}
\label{complex}
\end{figure}

\begin{figure}[t]
\epsfysize = 3.5cm
\centerline{\epsffile{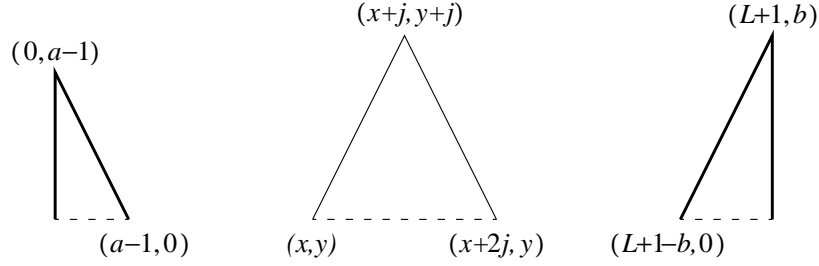}}
\caption{The graphical representation of pure particles.
The charges are, from left to right,
$(a-1)/2$, $j$ and $b/2$, respectively.}
\label{pure}
\end{figure}

We use the previous definition to define a very important
object, {\em a complex}.~\footnote{In ref.~\cite{Bressoud},
Bressoud has given a lattice path interpretation of the Andrews--Gordon
generalizations of the Rogers--Ramanujan
identities~\cite{Gordon,Andrews74}. In Bressoud's terminology
a complex corresponds to a {\em mountain}.}
This will be used subsequently to decompose each
restricted lattice path into particles.
\begin{definition}
A bulk complex is a lattice path from $(j,y_j)$ to $(k,y_k)$,
with $(j,y_j)$ and $(k,y_k)$ connected by
a dashed horizontal line,
such that $y_j=y_k$, $y_{\ell}>y_j$ for all $j<\ell<k$.

A left-boundary complex is a lattice path
from $(0,a-1)$ to $(j,0)$, such that
$y_k>0$ for all $k<j$, and with
$(0,0)$ and $(j,0)$ connected by a horizontal dashed line
and $(0,0)$ and $(0,a-1)$ connected by a vertical
solid line.

A right-boundary complex is a lattice path
from $(j,0)$ to $(L,b-1)$, $(L+1,b)$,
such that $y_k>0$ for all $k>j$ and with
$(L+1,0)$ and $(j,0)$ connected by a horizontal dashed line
and $(L+1,0)$ and $(L+1,b)$ connected by a vertical
solid line.
\end{definition}
Examples of a left-boundary, bulk and
right-boundary complex can be found in Fig.~\ref{complex}.
With respect to the above definition we remark that
the term complex is chosen since we
wish to view each complex as a collection of charged
particles moved on top of each other.
To make this explicit, we define particles
in the following two definitions.
\begin{definition}
A \underline{pure}
bulk particle of charge $j$ is a bulk complex with a single local
maximum of height $j$ (measured with respect to its dashed line).

A \underline{pure} left-boundary particle of charge $(a-1)/2$
is a left-boundary complex with a single local maximum,
located at $(0,a-1)$.

A \underline{pure} right-boundary particle of charge $b/2$
is a right-boundary complex with a single local maximum.
\end{definition}
The graphical representation of pure particles is given in
Fig.~\ref{pure}.

\begin{figure}[bt]
\epsfysize = 3.5cm
\centerline{\epsffile{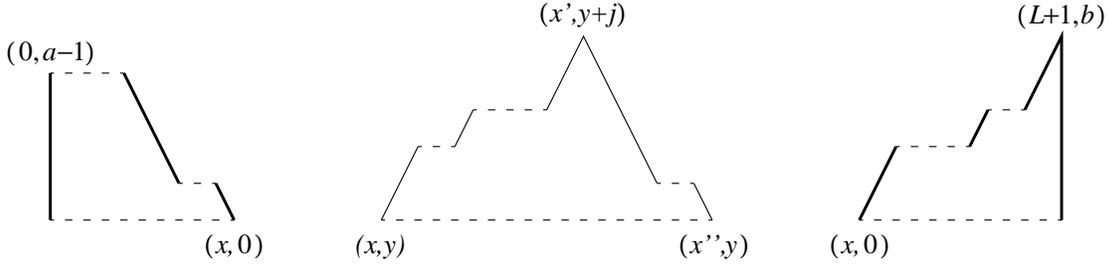}}
\caption{
Typical examples of a left-boundary, bulk
and right-boundary particle.
The charges of the particles are
$(a-1)/2$, $j$ and $b/2$, respectively.}
\label{unpure}
\end{figure}

To introduce the more general idea of a particle, we need some
simple terminology.
\begin{itemize}
\item The {\em peak} of a bulk complex is the
      left-most highest point. Similarly,
      the peak of a particle is its highest point.
\item The {\em origin} of a particle or complex
      is the left- and down-most point. \\
      The {\em endpoint} of a particle or complex
      is the right- and down-most point. \\
      The {\em baseline} of a particle or complex
      is the dashed line connecting the begin and
      endpoint.
\item The {\em contour} of a particle or complex
      is its part drawn with solid lines.
\end{itemize}
Using this we define
\begin{definition}
A bulk particle of charge $j$ is a pure bulk particle of charge $j$,
whose contour is interrupted at arbitrary integer points by horizontal
dashed lines of even length.

A left-boundary particle of charge $(a-1)/2$ is a pure
left-boundary particle of charge $(a-1)/2$, whose contour to the right
of $(0,a-1)$ is interupted at arbitrary integer points by horizontal
dashed lines of even length.

A right-boundary particle of charge $b/2$ is a pure
right-boundary particle of charge $b/2$, whose contour to the left
of $(L,b-1)$ is interupted at arbitrary integer points by horizontal
dashed lines of even length.
\end{definition}
Typical examples of particles are shown in Fig.~\ref{unpure}.
We note that for later convenience the contour of the boundary
particles is drawn with thicker lines than that
of the bulk particles.

\begin{figure}[t]
\epsfxsize = 10cm
\centerline{\epsffile{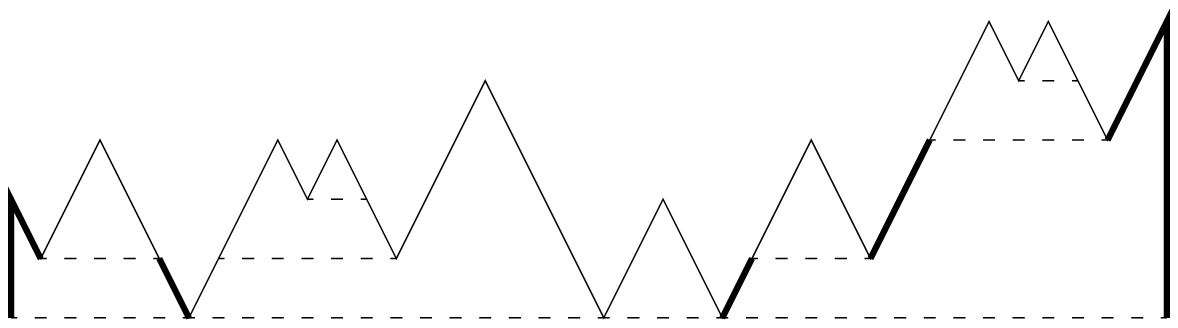}}
\caption{The particle configuration corresponding to the
restricted lattice path of Fig.~1.}
\label{content}
\end{figure}

With the above set of definitions we now give a prescription
to divide each restricted lattice path into particles.
This will be done by giving an algorithm that divides a complex
into a particle and several smaller complexes.
Each of these new complexes is either a particle
or is again divided into a particle
and yet smaller complexes.
This procedure is continued until the entire complex
is divided into particles. Since each lattice path can trivially
be divided into complexes, this gives a procedure to
divide any restricted lattice path into particles.
\begin{description}
\item[(0)]
Draw a dashed line along the $x$-axis from $(0,0)$ to $(L+1,0)$,
and draw bold lines from $(0,0)$ to $(0,a-1)$ and $(L+1,0)$
to $(L+1,b)$.
This divides each restricted lattice path into a
left-boundary complex, a right-boundary complex and a number of
bulk complexes.
For the restricted lattice path of Fig.~\ref{fig1}, we for
example get 4 complexes, 2 of which are of bulk-type.
If $a=1$, the left-boundary complex is absent.

Now consider each of the complexes obtained above. If such
a complex is a particle (in which case it is pure),
we are done with it.
If not, go to step (1) in case of a bulk complex and to
(1$_L$)  and (1$_R$) in case of a left- and
right-boundary complex, respectively.

\item[(1)]
Start at the peak of the complex and
move down to the right along the contour
till the endpoint of the complex.
When a local minimum is reached, i.e., the
contour starts going up again, we draw a dashed line
from this local minimum to the right until we cross the contour.
At that point we move further down along the contour.
If another minimum occurs we repeat the above, et cetera.

Repeat the above now moving to the left. That is,
start from the peak of the complex
and move down to the left till the origin of the complex.
If a local minimum is reached we
draw a dashed line to the left and continue our movement down when
the dashed line intersects the contour.

As a result of the above step we have divided the complex
into a particle (which is not pure)
and several (at least one) smaller complexes.
The peak and the baseline of the particle are the peak and
the baseline of the original complex.
Now go to (2).

\item[(1$_L$)]
Start from $(0,a-1)$.
Move to the right of this point down along the contour of the
complex till its endpoint.
If a local minimum is reached (which could be the point
$(0,a-1)$ itself), draw a dashed line
from this minimum to the right, until the contour is crossed.
At that point move further down along the
contour. If another minimum occurs we repeat the above, et cetera.

As a result of the above step we have divided the
left-boundary complex into a left boundary particle
and several (at least one) smaller bulk complexes.
To treat these smaller bulk complexes, go to (2).

\item[(1$_R$)]
Start from $(L+1,b)$.
Move to the left of this point down along the contour of the
complex till its endpoint.
If a local minimum is reached,
draw a dashed line from this minimum
to the left until the contour is crossed.
At that point move further down along the
contour. If another minimum occurs repeat the above, et cetera.

As a result of the above step we have divided the
right-boundary complex into a right-boundary particle
and several (at least one) smaller bulk complexes.
To treat these smaller bulk complexes, go to (2).

\item[(2)]
Scan each of the smaller bulk complexes. If such a
complex is a bulk particle (in which case it is pure),
we are done with it.
If not repeat step (1) for this complex.
\end{description}
We note that the above procedure converges, since the number of
local maxima of a restricted lattice path is finite.
In Fig.~\ref{content}, we have carried out the procedure
for the restricted lattice path of Fig.~\ref{fig1},
thereby identifying the corresponding configuration of
particles.\footnote{After having identified all particles, we
implicitly assume the step of (re)drawing the contour of the
boundary particles with fat lines.}

Thanks to the above algorithm, each restricted lattice path
in rlp$(0,r)$ can now be viewed as a particle configuration.
In particular, since the maximal height of a path is
$r-2$, we have bulk particles of charge 1 up to $r-2$, as well
as a left-boundary particle of charge $(a-1)/2$ and a
right-boundary particle of charge $b/2$.
The contour of a bulk particle of charge $j$ consists
of $j$ up and $j$ down steps, the contour of a
left-boundary particle of $a-1$
down steps and the contour of a
right-boundary particle of $b$
up steps.
Letting $n_j$ denote the number of bulk particles
of charge $j$, we thus have the completeness relation
\begin{equation}
a + b - 1 + 2\sum_{j=1}^{r-2} j\, n_j = L+1.
\label{completeness}
\end{equation}
Using this relation,
$n_{r-2}$ can be computed given the occupation numbers
$n_1,\ldots,n_{r-3}$. For this reason (and anticipating
things to come), we define the column
vector $\vec{n}= \,^T(n_1,\ldots,n_{r-3})$,
and when we say
``a restricted lattice path has particle content $C=\vec{n}\,$'',
we mean by this the particle content $C=\{n_1,\ldots,n_{r-2}\}$
subject to the restriction (\ref{completeness}).

Having associated a configuration of
particles with each path in rlp$(0,r)$,
we define rlp$(\vec{n})$ as the subset of
paths in rlp$(0,r)$, with particle content
$\vec{n}$.
This puts us in a position to properly define what
we mean by the Fermi-gas partition function as
introduced in (\ref{ZCab}),
\begin{equation}
Z(C;a,b)  = Z(\vec{n};a,b) =
\sum_{\mbox{\scriptsize rlp}(\vec{n})}
\e^{\; \displaystyle -\beta \sum_{j=1}^L E(j)},
\end{equation}
with energy function defined in (\ref{energy}).

So far, we have repeatedly used the term
Fermi-gas, without any clear motivation.
Clearly, we have defined all allowed
configurations of our one-dimensional system of charged
particles, as well its Hamiltonian or energy
function, but the actual nature of the system remains
rather elusive.
However, in our actual computation of $Z$, in the next
subsection, it turns out to be expedient to
define rules of motion that allow one to
obtain any configuration with content $\vec{n}$
from a given so-called minimal configuration
with the same content.
These rules of motion have a clear fermionic character, in
that particles of the same charge cannot exchange
position, unlike particles of different charge.

\subsection{Computation of $Z(\vec{n};a,b)$.}
In this section we compute the partition function
of the one-dimensional Fermi-gas.
Throughout the section we assume the
particle content to be $\vec{n}$.

\begin{figure}[t]
\epsfxsize = 11cm
\centerline{\epsffile{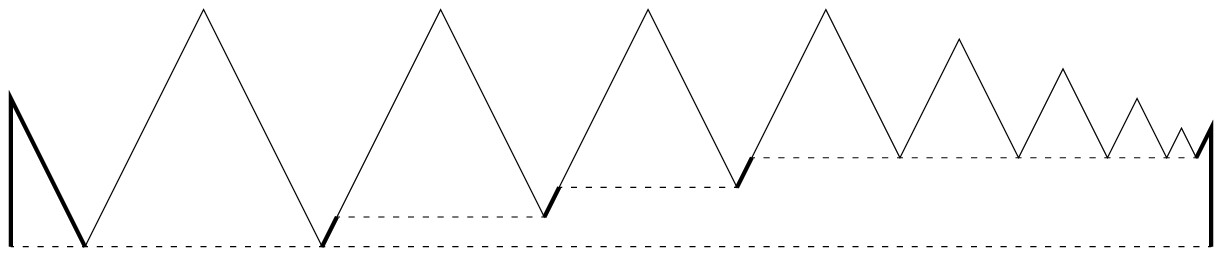}}
\caption{The restricted lattice path corresponding to the
minimal configuration of particles.
Here each bulk particle of charge $j$
has to be copied $n_j$ times. The dashed lines are the
baselines of the particles.}
\label{min}
\end{figure}

To compute the sum over all particle configurations,
we first select a particular configuration called
the {\em minimal configuration}.\footnote{From a statistical
mechanics point of view {\em ground state configuration}
may be more appropriate, but we prefer
to conform to our earlier naming in ref.~\cite{W}.}
It will be defined purely graphically.
\begin{definition}
The configuration shown in Fig.~\ref{min} is called the
minimal configuration. Here each bulk particle of charge
$j$ should be repeated $n_j$ times, i.e.,
$$
\centerline{\epsffile{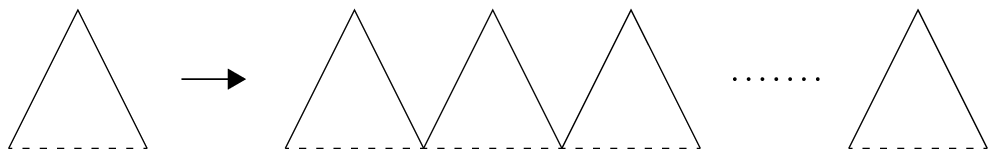}}
$$
\end{definition}
Note that in the minimal configuration
\begin{itemize}
\item All (bulk) particles are positioned as much to the right and up
as possible, the baseline of the particles of charge $j$ having
$y$-coordinate equal to $\min(b-1,r-j-2)$.
\item The particles are positioned
      in order of decreasing charge.
\item Apart from the right-boundary particle, all particles are pure.
\end{itemize}

\subsubsection{Contribution of the minimal configuration}
To compute the weight of the minimal configuration, we use that the
energy $E_j(x)$ of a pure bulk particle
of charge $j$, with origin at position $x$
and endpoint at position $x+2j$,
is given by
\begin{equation}
E_j(x)=
\frac{1}{2} \; \sum_{
{k=1 \atop k\neq j}}^{2j-1} (k+x)
= (j+x)(j-1).
\label{Ejx}
\end{equation}
Similarly, we get for the energy $E_a$ of the pure
left-boundary particle
with charge $(a-1)/2$,
\begin{equation}
E_a=\frac{1}{4}(a-1)(a-2).
\label{Ea}
\end{equation}
A bit more work is required to obtain the energy $E_b$ of the
right-boundary particle with charge $b$, since its contour
is broken into $b$ segments all of length 1.
Summing up the $b$ different contributions leads to
\begin{equation}
E_b=\frac{1}{4}(b-1)(2a+b-2) + \sum_{j=r-b}^{r-2}
j(b-r+j+1) n_j.
\end{equation}
Using the above three  results, we compute the energy of
the minimal configuration as
\begin{eqnarray}
E_{\min}
&=& E_a+E_b+\sum_{j=1}^{r-2} \sum_{\ell=1}^{n_j}
E_j\biggl(a-2+\min(b,r-j-1)+2j(\ell-1)
+ 2\sum_{k=j+1}^{r-2} k \, n_k \biggr)
\nonumber \\
&=& \sum_{j=1}^{r-2} (j-1) n_j \biggl( j \, n_j +
2 \sum_{k=j+1}^{r-2} k \, n_k \biggr) +
(a+b-2) \sum_{j=1}^{r-2}(j-1) n_j \\
& & + \sum_{j=r-b}^{r-2} (b-r+j+1) n_j
+ \frac{1}{4}(a-1)(a-2) + \frac{1}{4}(b-1)(2a+b-2).
\nonumber
\end{eqnarray}
To simplify this expression, we eliminate $n_{r-3}$ using the
completeness relation (\ref{completeness}). This yields
\begin{eqnarray}
\lefteqn{
E_{\min} =
\sum_{j=1}^{r-3}
\left(
\sum_{k=1}^j \frac{k (r-j-2)}{r-2} +
\sum_{k=j+1}^{r-3} \frac{j (r-k-2)}{r-2}
\right)
n_j \, n_k }
\nonumber \\
& & \qquad
- \left(
\sum_{j=1}^{r-b-1} \frac{j (b-1)}{r-2} +
\sum_{j=r-b}^{r-3} \frac{(r-b-1)(r-j-2)}{r-2}
+ L \sum_{j=1}^{r-3} \frac{r-j-2}{r-2} \right) n_j \\
& & \qquad
+ \: \frac{L^2(r-3) + 2 L (b-1) -(a-1)(r-a-1)+(b-1)(r-b-1)}{4(r-2)}.
\nonumber
\end{eqnarray}
We now recall the definition of the inverse Cartan matrix of
the Lie algebra A$_{r-3}$,
\begin{equation}
\renewcommand{\arraystretch}{2.4}
C^{-1}_{j,k} =
\left\{
\begin{array}{cc}
\displaystyle
\frac{k(r-j-2)}{r-2} \qquad & k \leq j \\
\displaystyle
\frac{j(r-k-2)}{r-2} & k \geq j .
\end{array}
\right.
\label{invC}
\end{equation}
Using this, we finally  obtain
\begin{lemma}\label{lemmin}
The energy of the minimal configuration is given by
\begin{eqnarray}
\lefteqn{
E_{\min}
=\sum_{j,k=1}^{r-3}
\left(n_j-\frac{L}{2} \, \delta_{j,1}-\frac{1}{2}\, \delta_{j,r-b-1}
-\frac{1}{2}\, \delta_{j,r-a-1} \right)
C^{-1}_{j,k} } \nonumber \\
& & \qquad \qquad\qquad \qquad \qquad \times
\left(n_k-\frac{L}{2}\, \delta_{k,1}-\frac{1}{2}\, \delta_{k,r-b-1}
+\frac{1}{2}\, \delta_{k,r-a-1} \right).
\label{Emin}
\end{eqnarray}
\end{lemma}

\subsubsection{Contribution of the non-minimal configurations}
\begin{figure}[hbt]
\epsfxsize = 13cm
\centerline{\epsffile{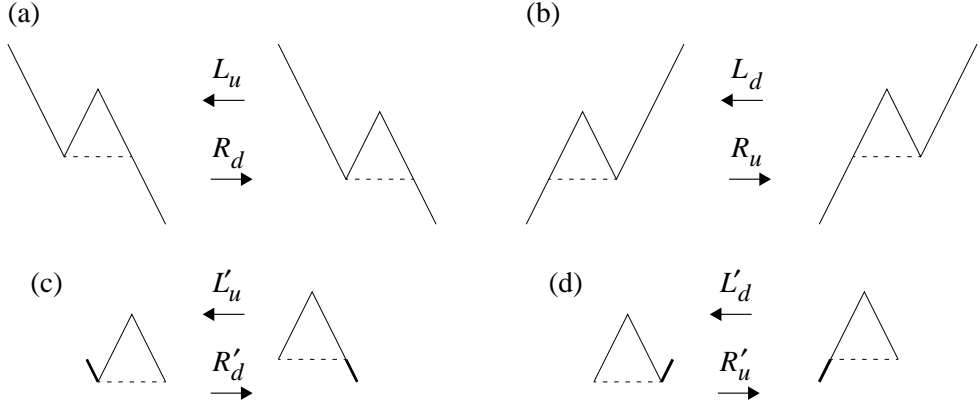}}
\caption{
(a) The moves $L_u$ and $R_u$.
(b) The moves $L_d$ and $R_u$.
(c) The moves $L_u'$ and $R_u'$.
(d) The moves $L_d'$ and $R_u'$.}
\label{moves}
\end{figure}

To compute the contribution to the partition function of the
other configurations, we define rules of motion
which generate all non-minimal configurations from the minimal one.
These rules break up into several different {\em
elementary moves} as follows.
\begin{definition}\label{defmoves}
Let $X=\{(x_1,y_1),(x_2,y_2),(x_3,y_3),(x_4,y_4)\}$
denote a sequence of four points on the contour of a
configuration, each pair of consecutive points connected
straight lines, such that the contour
in between $(x_1,y_1)$ and $(x_4,y_4)$ does not
belong to a boundary particle.
We may then replace this sequence by a new sequence
of four points as follows.
\begin{description}
\item[move $L_u$:]
If $y_4\leq y_2 <y_3<y_1$,

\noindent
$\qquad \qquad \qquad \quad
L_u(X)=
\{(x_1,y_1),(x_2-1,y_2+1),(x_3-1,y_3+1),(x_4,y_4)\}$.
\item[move $R_d$:]
If $y_4<y_2<y_3\leq y_1$,

\noindent
$\qquad \qquad \qquad \quad
R_d(X)=
\{(x_1,y_1),(x_2+1,y_2-1),(x_3+1,y_3-1),(x_4,y_4) \}$.
\item[move $L_d$:]
If $y_1< y_3 <y_2\leq y_4$,

\noindent
$\qquad \qquad \qquad \quad
L_d(X)=
\{(x_1,y_1),(x_2-1,y_2-1),(x_3-1,y_3-1),(x_4,y_4)\}$.
\item[move $R_u$:]
If $y_1\leq y_3 <y_2< y_4$,

\noindent
$\qquad \qquad \qquad \quad
R_u(X)=
\{(x_1,y_1),(x_2+1,y_2+1),(x_3+1,y_3+1),(x_4,y_4) \}$.
\end{description}
\end{definition}
Besides these ``bulk-type'' moves we need some special
boundary moves.
\begin{definition}\label{defmovesl}
Let $X=\{(x_1,y_1),(x_2,y_2),(x_3,y_3),(x_4,y_4)\}$
be four points on the contour of a configuration,
each pair of consecutive points connected
by a straight line.
We may then replace $X$ as follows.
\begin{description}
\item[move $L'_u$:]
Let $(x_1,y_1)=(x_2-1,y_2+1)$. If $y_2=y_4<r-2$ and
the contour between the first two points belongs to the
left-boundary particle,

\noindent
$\qquad \qquad
L'_u(X)=
\{(x_2-1,y_2+1),(x_3-1,y_3+1),(x_4-1,y_4+1),(x_4,y_4)\}$,

\noindent
where the contour between the last two points belongs to the
left-boundary particle.
\item[move $R'_d$:]
Let $(x_4,y_4)=(x_3+1,y_3-1)$. If $y_1=y_3<2$
and the contour between the last two points belongs to the
left-boundary particle,

\noindent
$\qquad \qquad
R'_d(X)=
\{(x_1,y_1),(x_1+1,y_1-1),(x_2+1,y_2-1),(x_3+1,y_3-1)\}$,

\noindent
where the contour between the last two points belongs to the
left-boundary particle.
\item[move $L'_d$:]
Let $(x_1,y_1)=(x_2-1,y_2-1)$. If $y_2=y_4<y_3$
and the contour between the first two points belongs to the
right-boundary particle,

\noindent
$\qquad \qquad
L'_d(X)=
\{(x_2-1,y_2-1),(x_3-1,y_3-1),(x_4-1,y_4-1),(x_4,y_4)\}$,

\noindent
where the contour between the last two points belongs to the
right-boundary particle.
\item[move $R'_u$:]
Let $(x_4,y_4)=(x_3+1,y_3+1)$. If $y_1=y_3<y_2<r-2$,
$y_3<b-1$
and the contour between the last two points belongs to the
right-boundary particle,

\noindent
$\qquad \qquad
R'_u(X)=
\{(x_1,y_1),(x_1+1,y_1+1),(x_2+,y_2+1),(x_3+1,y_3+1)\}$,

\noindent
where the contour between the first two points belongs to the
right-boundary particle.
\end{description}
\end{definition}
For the graphical interpretation of this long list of moves,
see Fig.~\ref{moves}.

To fully appreciate these moves, we list its main characteristics in
several lemmas, which are at the core of our fermionic computation
of the one-dimensional configuration sums.

\begin{lemma}
The elementary moves are reversible. That is,
if there is a move of type $M_s^p$ from a configuration $C$
to a configuration $C'$, then there is a move of type
$\bar{M}_{\bar{s}}^p$
from $C'$ to $C$.
Here $M=L$ or $R$, $s=u$ or $d$, $p=\quad,\, '$ or $''$ and
$\bar{R}=L$, $\bar{L}=R$, $\bar{u}=d$ and $\bar{d}=u$.
\end{lemma}
Proof: Let us show this for $L_u$. The other moves follow in
similar manner.
Let $X$ be a sequence of
four extrema as in definition~\ref{defmoves}, satisfying
$y_4\leq y_2 <y_3<y_1$. Hence we can carry out $L_u$ to obtain
$X'=L_u(X)=\{(x'_1,y'_1),(x'_2,y'_2),(x'_3,y'_3),(x'_4,y'_4)\}$.
{}From the definition of the move $L_u$,
we find that $y_4'\leq y_2'-1 < y_3'-1<y_1$. We rewrite this to obtain
$y_4'< y_2' < y_3'\leq y_1$ and hence we can carry out the
move $R_d$ to obtain $R_d(X')=X$.
$\Box$

\begin{lemma}\label{lem2}
The moves leave the particle content $\vec{n}$ fixed.
\end{lemma}
Proof: This follows immediately from the graphical representation
of the moves shown in Fig.~\ref{moves}, where the dashed lines
represent the baselines of the pure particles being moved.
Note here that the graphical representations of the
moves $R_u$ and $L_d$ are the generic cases.
Performing a move of type $R_u$ to a sequence $X$ as defined in
definition~\ref{defmoves}, with $y_2=y_4-1$, may lead to a ``jump''
of the baseline.
A similar thing may happen when performing a move of type $L_d$
to a sequence with $y_2= y_4$:
$$
\hspace{-5 cm}
\epsfxsize = 6	cm
\centerline{\epsffile{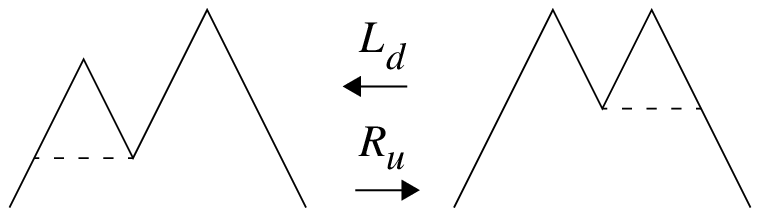}}
\hspace{-4 cm}
\Box$$

\begin{lemma}
Given the minimal configuration, we cannot
make any of the $R$-type moves.
\end{lemma}
Proof:
We can only make moves of type $R_d$ if we have
a sequence of $X$ as in definition~\ref{defmoves}, with
$y_4<y_2$. Clearly this does not occur.
We can only make moves of type $R_u$ if we have
a sequence $X$, with $y_2 < y_4$.
Again this does not occur.
We cannot make a move of type $R'_d$ since the
left-boundary particle is in its pure form.
Finally, we cannot make a move of type $R'_u$
since all particles of charge $j\geq r-b-1$ have their peak
at $y=r-2$, and all particles of charge $j\leq r-b-1$ have their
endpoint at $y=b-1$. $\Box$

\begin{lemma}
If a configuration is not the minimal one, we can always
make a move of type $R$.
\end{lemma}
Proof:
By construction the minimal configuration is the only configuration
that does not meet any of the conditions required for one of the
$R$-type moves. In particular, all maxima (apart from the
initial point of the path) are of decreasing order and all
minima of increasing order. This completely fixes the path.
If one of these two properties is broken somewhere along the
path, we can always make an $R$-type move. $\Box$

These first four lemmas can be combined to give the
following proposition:
\begin{proposition}
All non-minimal paths are generated by moves of type $L$
from the minimal configuration.
All non-minimal configurations can be reduced to the
minimal configuration by moves of type $R$.
\end{proposition}

Having established the above proposition, we can perform the actual
calculation of the generation function $\cal C$
of the moves of type $L$.
Again we prepare some lemmas to obtain the desired result.
\begin{lemma}\label{genqq}
Each move of type $L$ generates a factor $q$.
\end{lemma}
Proof: We show this for the typical case of move $L_u$.
The total energy $E$ of a sequence of extrema $X$ is
\begin{equation}
E=\sum_{{j=x_1+1 \atop j\neq x_2,x_3}}^{x_4-1} j.
\end{equation}
Similarly, the energy $E'$ of the sequence $X'=L_u(X)$ is
\begin{equation}
E=\sum_{{j=x_1+1 \atop j\neq x_2-1,x_3-1}}^{x_4-1} j.
\end{equation}
Hence we find
\begin{equation}
\e^{-\beta (E'-E)} = q^{E'-E} = 1. \qquad \quad \Box
\end{equation}

In the following it will be convenient to label the
bulk particles in the minimal configuration, letting
$p_{j,\ell}$ denote the $\ell$-th particle of charge $j$,
counted from the left.
To now generate all non-minimal configurations, we
give an ordering for carrying out the moves of type $L$.
\begin{itemize}
\item
The particle $p_{j,\ell}$ is moved to the left using
moves of type $L$, prior to any of the particles
$p_{k,m}$, with $k\leq j$, and with $m>\ell$ if $k=j$.
\end{itemize}

Assuming this order (which will be justified later), we have
\begin{lemma}\label{lemmj}
The maximal number of $L$-type moves $p_{j,1}$ can make is
\begin{equation}
m_j=2 \sum_{k=j+1}^{r-2} (k-j) \, n_k
+\min(a-1,r-j-2) + \min(b-1,r-j-2).
\label{mj}
\end{equation}
\end{lemma}
Proof:
We proof this lemma in two steps. In the first step
(\ref{mj}) is shown to be true
for the minimal configuration,
and in the second step it is shown that $m_j$ is invariant under
having moved the particles $p_{k,m}$, with $k>j$, prior
to $p_{j,1}$.

Let us start to
calculate the number of $L$-type moves
needed to exchange the position of two particles of charge
$k$ and $j$, $k>j$, with $j$ positioned immediately
to the right of $k$.
In such a configuration of two particles we have a
sequence $X=\{(x_1,y_1),\ldots,(x_5,y_5)\}$ of
points connected by straight lines,
with $y_1=y_3=y_5$ and $y_2=k$ and $y_4=j$.
{}From these conditions it follows that move $L_u$
can be carried out $k-j$ times
to the sequence $\{(x_2,y_2),\ldots,(x_5,y_5)\}$.
This gives a new sequence
$X'=\{(x'_1,y'_1),\ldots,(x'_5,y'_5)\}$,
with $y'_1=y'_5$, $y'_2=y'_4=k$ and $y'_3=k-j$.
{}From these conditions it follows that move $L_d$
can be carried out $k-j$ times
to the sequence $\{(x'_1,y'_1),\ldots,(x'_4,y'_4)\}$.
This gives the final sequence $X''=
\{(x''_1,y''_1),\ldots,(x''_5,y''_5)\}$,
with
$y^{''}_1=y^{''}_3=y^{''}_5$, $y^{''}_2=j$ and $y^{''}_4=k$.
The total number of moves carried out is therefore $2(k-j)$.
Since in the minimal configuration
there are $n_k$ particles of charge $k$
to the left of $p_{j,1}$, this gives a total contribution
$\sum_{k=j+1}^{r-2} (k-j) \, n_k$.
Apart from this, we encounter
the situation where immediately to the left of $p_{j,1}$
we have a segment of the right-boundary particle.
In such an instant we can perform $L'_d$, moving
$p_{j,1}$ one step down. By construction of the
minimal configuration, this occurs
$\min(b-1,r-j-2)$ times.
Finally,  after having descended all the way down and
having exchanged position with all particles of charge
$>j$, $p_{j,1}$ is positioned immediately to the right
of the left-boundary particle. It can then move up
exactly $\min(a-1,r-j-2)$ times using move $L'_u$.
Adding up all the contributions gives (\ref{mj}).

To see that (\ref{mj}) is unaltered by first having moved
some (or all) particles of charge greater than $j$,
consider a sequence of four points
$X=\{(x_1,y_1), (x_2,y_2), (x_3,y_3), (x_4,y_4)\}$
connected by straight lines.
First, let $y_1>y_2<y_3>y_4$ and let $p_{j,1}$ be positioned
immediately to the right of the sequence, i.e., the
origin of $p_{j,1}$ is at $(x_4,y_4)$. Also, let the
contour between the first two points not belong to the
left-boundary particle.
The total number of $L$-type steps
$p_{j,1}$ can make is then
$(y_3-y_4-j)+(y_3-y_2-j)+(y_1-y_2-j)=x_4-x_1-3j$, which is
independent of the positions of the points $(x_2,y_2)$ and
$(x_3,y_3)$. Hence carrying out any moves to $X$ does not
change the number of moves $p_{j,1}$ can make relative to $X$.
If the contour between the first two point does belong to the
left-boundary particle, this is changed to
$x_4-x_1-3j +r-2-\min(r-2-j,y_1)$ which is still independent of
the relative positions of $(x_2,y_2)$ and $(x_3,y_3)$.
Second, let $y_1<y_2>y_3<y_4$ and let $p_{j,1}$ be positioned
immediately to the left	of the sequence, i.e., the
endpoint of $p_{j,1}$ is at $(x_1,y_1)$. Also, let the
contour between the last two points not belong to the
left-boundary particle.
The total number of $R$-type steps
$p_{j,1}$ can make is then
$(y_2-y_1-j)+(y_2-y_3-j)+(y_4-y_3-j)=x_4-x_1-3j$, which is
independent of the positions of the points $(x_2,y_2)$ and
$(x_3,y_3)$.  Thanks to reversibility, the number of $L$-type
moves $p_{j,1}$ can make relative to $X$ is also $x_4-x_3-3j$.
If the contour between the last two point does belong to the
right-boundary particle, this again chances by a term
independent of the detailed positions of $(x_2,y_2)$ and $(x_3,y_3)$.
$\Box$
\begin{lemma}\label{lemkj}
The maximal number of $L$-type moves $p_{j,\ell}$ can make
is $k_{j,\ell-1}$, with $k_{j,\ell-1}$ the actual number
of steps taken by $p_{j,\ell-1}$
\end{lemma}
At last!, we finally encountered the fermionic nature
of our lattice-gas.
Proof:
Assume $p_{j,\ell-1}$ has made $k_{j,\ell-1}$ moves.
Obviously, (before) the first $k_{j,\ell-1}$ moves,
$p_{j,\ell}$ ``sees'' the same contour immediately
to its left as $p_{j,\ell-1}$ did, when carrying out
its leftward motion. Since
$p_{j,\ell-1}$ and $p_{j,\ell}$ are identical particles,
$p_{j,\ell}$ can thus carry out at least $k_{j,\ell-1}$ moves.
Let $p_{j,\ell}$ indeed carry out $k_{j,\ell-1}$ moves.
After that we encounter the situation of
two pure particles of charge $j$, with endpoint of
the first being origin of the next.
The right-most of the two can neither
carry out $L_u$, nor $L_d$, since
(in the notation of definition~\ref{defmoves})
$y_1=y_3$.
$\Box$

We note that the above two lemmas justify the chosen
ordering of carrying out the leftward moves.
First of all, by lemma~\ref{lemkj} it follows that
we indeed have to move $p_{j,\ell-1}$ before $p_{j,\ell}$.
Furthermore, we have to move $p_{k,m}$ before $p_{j,\ell}$,
$k>j$  since the elementary moves
only allow for leftward motion of pure particles,
see Fig.~\ref{moves}. Finally we have seen in the proof
of lemma~\ref{lemmj} that the number of moves the particles
of charge $j$ can make is independent of the actual
configuration of particles of charge $>j$.

\begin{lemma}
The contribution to the generating function $C$
of the particles of charge $j$, is given by
$\cal C$, is given by
\begin{equation}
{\cal C}_j = \Mult{m_j+n_j}{n_j}.
\label{Cj}
\end{equation}
\end{lemma}
Proof:
{}From the lemmas~\ref{genqq}, \ref{lemmj} and \ref{lemkj} we get
(dropping the subscripts $j$ in the $k$-variables)
\begin{equation}
{\cal C}_j=
\sum_{k_1=0}^{m_j} \sum_{k_2=0}^{k_1} \ldots
\sum_{k_{n_j}=0}^{k_{n_j-1}} q^{k_1+k_2+\cdots +
k_{n_j}}.
\end{equation}
We can (re)interpret this sum as the generating
function of all partitions with largest part less or
equal to $m_j$
and number of parts less or equal to $n_j$. Thus we get
(\ref{Cj}), see e.g., ref.~\cite{Andrews}.

Combining the above lemma with lemma~\ref{lemmin},
we can state our second proposition as
\begin{proposition}
The partition function of the one-dimensional Fermi-gas is
given by
\begin{equation}
Z(\vec{n};a,b)=
q^{E_{\min}} \prod_{j=1}^{r-3} {\cal C}_j=
q^{E_{\min}} \prod_{j=1}^{r-3}
\Mult{m_j + n_j}{n_j},
\label{coll}
\end{equation}
with $E_{\min}$ given by (\ref{Emin})
and $m_j$ by (\ref{mj}).
\end{proposition}
To recast this result into a simpler from, we
eliminate the $n$-variables in favour of the $m$-variables.
To do so we use the simple formulae
\begin{eqnarray}
\lefteqn{
-\min(p,q-1)+2 \,\min(p,q)-\min(p,q+1)=\delta_{p,q}
\qquad \: p,q-1 \geq 0 } \nonumber \\
\lefteqn{
-\min(p,q-1)+2 \, \min(p,q)=p+\delta_{p,q}
\qquad \qquad \quad  \qquad 0\leq p \leq q+1  }
\end{eqnarray}
to get
\begin{equation}
-m_{j-1}+2m_j-m_{j+1}=
L \, \delta_{j,1} + \delta_{j,r-a-1} +
\delta_{j,r-b-1} -2 n_j
\qquad  j=1,\ldots,r-3
\label{mn}
\end{equation}
with $m_0=m_{r-2}=0$. To obtain the $j=1$ case
of the above equation we made use of the completeness
relation (\ref{completeness}).
Introducing the $(r-3)$-dimensional vectors $\vec{m}$ and
$\vec{\e}_j$ with entries $(\vec{m})_j=m_j$
and $(\vec{\e}_j)_k=\delta_{j,k}$,
we can rewrite (\ref{mn}) as
\begin{equation}
\vec{n} = \frac{1}{2}\: (L \, \vec{\e}_1 + \vec{\e}_{r-a-1} +
\vec{\e}_{r-b-1} - C \: \vec{m}).
\label{mnsystem}
\end{equation}
Substituting this into equations~(\ref{Emin}) and (\ref{coll}),
we arrive at the following simple result:
\begin{proposition}
The partition function of the Fermi-gas of content $\vec{n}$
reads
\rm
\begin{equation}
Z(\vec{n};a,b)  =
q^{\, \case{1}{4} \,  \vec{m}^T C \, \vec{m}
-\case{1}{2} \,m_{r-a-1} }
\prod_{j=1}^{r-3}
\Mult{\frac{1}{2}({\cal I} \, \vec{m} + L \, \vec{\e}_1 +
\vec{\e}_{r-a-1} +\vec{\e}_{r-a-1})_j}{m_j},
\label{Z}
\end{equation}
\em
whth, $\vec{m}$ obtained through equation~(\ref{mnsystem}).
\end{proposition}
\subsection{Computation of $\Xi_L(a,b)$.}
Having computed the partition function of our
Fermi-gas, it is only a trivial
step to obtain the grand-canonical partition function
$\Xi_L(a,b)$, defined in (\ref{ZCab}). In particular
\begin{equation}
\Xi_L(a,b) = \sum_{\vec{n}} Z(\vec{n};a,b) .
\end{equation}
Since our final result (\ref{Z}) for $Z$ is entirely expressed
through the $m$-variables, it is natural to also
express the above sum over $\vec{n}$ in terms of a sum over
$\vec{m}$.
{}From $(\ref{mj})$, and the fact that the occupation numbers $n_j$
cannot be negative, we get
\begin{equation}
m_j = \min(a-1,r-j-2) + \min(b-1,r-j-2) + 2 \Integer_{\geq 0}
\qquad j=1,\ldots,r-3.
\label{rm}
\end{equation}
Hence we obtain the grand-canonical partition function as
\begin{equation}
\Xi_L(a,b) =
\left. \sum_{\vec{m}}
\right.^{(0)}
q^{\, \case{1}{4} \,  \vec{m}^T C \, \vec{m}
-\case{1}{2} \,m_{r-a-1} }
\prod_{j=1}^{r-3}
\Mult{\frac{1}{2}({\cal I} \, \vec{m} + L \, \vec{\e}_1 +
\vec{\e}_{r-a-1}
+\vec{\e}_{r-a-1}
)_j}{m_j},
\label{Xi}
\end{equation}
where the $(0)$ in the sum over $\vec{m}$ denotes the restriction
(\ref{rm}).

\subsection{Computation of $X_L(a,b)$.}
To finally obtain the one-dimensional configuration sum
$X_L(a,b)$, we have to carry out the sum (\ref{summu}),
where we recall that $\Xi_L(a,b):=\Xi_L(a,b,0,r)$.

To get the expression for $\Xi_L(a-\mu,b-\mu,0,r-\mu)$,
we have to make the substitutions
$a\to a-\mu$, $b\to b-\mu$ and $r\to r-\mu$ in (\ref{Xi}).
This exactly gives back (\ref{Xi}) apart from the fact that
the restriction on the sum changes to
\begin{equation}
m_j = \left\{
\begin{array}{ll}
\min(a-1,r-j-2) & \\
\quad + \min(b-1,r-j-2)-2\mu + 2 \Integer_{\geq 0} \qquad &
j=1,\ldots,r-\mu-3 \\
0  & j=r-\mu-2,\ldots,r-3.
\end{array} \right.
\label{rmmu}
\end{equation}
Denoting this restriction as $(\mu)$, we can write
\begin{equation}
X_L(a,b)=\sum_{\mu=0}^{\min(a,b)-1}
\left. \sum_{\vec{m}}
\right.^{(\mu)}
q^{\,\case{1}{4} \,  \vec{m}^T C \, \vec{m}
-\case{1}{2} \,m_{r-a-1} }
\prod_{j=1}^{r-3}
\Mult{\frac{1}{2}({\cal I} \, \vec{m} + L \, \vec{\e}_1 +
\vec{\e}_{r-a-1} +\vec{\e}_{r-a-1})_j}{m_j}.
\label{ds}
\end{equation}
Combining the sum over $\vec{m}$ restricted to $(\mu)$ and the
sum over $\mu$, gives
\begin{equation}
X_L(a,b)=
\left. \sum_{\vec{m}}
\right.^{'}
q^{\, \case{1}{4} \,  \vec{m}^T C \, \vec{m}
-\case{1}{2} \,m_{r-a-1} }
\prod_{j=1}^{r-3}
\Mult{\frac{1}{2}({\cal I} \, \vec{m} + L \, \vec{\e}_1 +
\vec{\e}_{r-a-1} +\vec{\e}_{r-a-1})_j}{m_j},
\label{ss}
\end{equation}
with the prime denoting yet another restriction,
\begin{equation}
m_j \equiv \min(a-1,r-j-2) + \min(b-1,r-j-2)
\qquad j=1,\ldots,r-3.
\label{rmprime}
\end{equation}
Unfortunately, we have not found an elegant way to
prove this simplification and we defer it till the
appendix.

To rewrite the above form of the restriction,
in the form conjectured
in refs.~\cite{KKMMa,Melzer},
we note the identity
\begin{equation}
\begin{array}{rcccccccc}
\min(p,q) & \equiv & \delta_{p+1,q}&+&\delta_{p+3,q}&+&
\delta_{p+5,q} & + & \ldots \\
&+ & \delta_{1,q}&+&\delta_{3,q}&+&\delta_{5,q}& + &\ldots \; ,
\end{array}
\end{equation}
for $p,q\geq 0$.
Using this twice, once setting
setting $p=a-1$ and $q=r-j-2$, and once setting
$p=b-1$ and $q=r-j-2$, we get
$m_j \equiv (\vec{Q}_{a,b})_j$, with $\vec{Q}_{a,b}$ given by
(\ref{Qrest}).

We can thus
conclude this section formulating our main result as a theorem.
\begin{theorem}\label{pageth1}
\renewcommand{\arraystretch}{1}
For all $a=1,\ldots,r-1$, $b=1,\ldots,r-2$ and
$L -|a-b| \in  2 \, \Integer_{\geq 0}$,
the one-dimensional configuration sum
(\ref{confsums}), is given by
\rm
\begin{equation}
X_L(a,b) =
\sum_{\vec{m} \equiv \vec{Q}_{a,b} }
q^{\, \case{1}{4} \,
\vec{m}^T C \, \vec{m}
-\case{1}{2} \,m_{r-a-1} }
\prod_{j=1}^{r-3}
\Mult{\frac{1}{2}({\cal I} \, \vec{m} + L \, \vec{\e}_1 +
\vec{\e}_{r-a-1} + \vec{\e}_{r-b-1})_j}{m_j},
\nonumber
\label{Xferm}
\end{equation}
\em
where  $\vec{Q}_{a,b}$ is given by (\ref{Qrest}).
\end{theorem}

\nsection{Bosonic solution of the ABF model}\label{sec4}
In this section we recall the method for computing
the sum (\ref{confsums}) to obtain (up to a prefactor)
the right-hand side of Melzer's identities (\ref{Mid}).
This alternative  approach to the sum (\ref{confsums}) is
the one originally taken by Andrews, Baxter and Forrester~\cite{ABF}
and is given here mainly for reasons of completeness.

As a first step we introduce a function $Y_L(a,b)$
defined exactly as $X_L(a,b)$ in (\ref{confsums}), but
with $\sigma_{L+1}=b-2$ instead of $\sigma_{L+1}=b$.
We can then immediately infer the
recurrence relations
\begin{eqnarray}
X_L(a,b) &=& \, Y_{L-1}(a,b+1) + q^{L/2} X_{L-1}(a,b-1)
\quad \qquad  1 \leq b \leq r-2  \label{rr1} \\
Y_L(a,b) &=& X_{L-1}(a,b-1) + q^{L/2} \, Y_{L-1}(a,b+1)
\quad \qquad  2 \leq b \leq r-1,
\label{rr2}
\end{eqnarray}
subject to the initial and boundary conditions
\begin{eqnarray}
\lefteqn{
X_0(a,b) = Y_0(a,b)=\delta_{a,b}} \label{init}\\
\lefteqn{X_L(a,0)=Y_L(a,r)=0.}
\label{bound}
\end{eqnarray}
To state the solution to these equations, we quote
the following theorem established by Andrews, Baxter and
Forrester~\cite{ABF}:

\begin{theorem}
For $L\geq 0$, $1 \leq a,b,c \leq r-1$, $c=b\pm 1$, $L+a-b \equiv 0$,
let $X_L(a,b,c) := X_L(a,b)$ if $c=b+1$ and
$X_L(a,b,c)  := Y_L(a,b)$ if $c=b-1$. Then
\begin{eqnarray}
\lefteqn{
X_L(a,b,c) = q^{(a-b)(a-c)/4} \sum_{j=-\infty}^{\infty} \left\{
q^{j\bigl(r(r-1)j+r(b+c-1)/2-(r-1)a\bigr)}
\Mults{L}{\frac{1}{2}(L+a-b)-rj}\right. }
\nonumber \\
& & \left. \qquad \qquad \qquad \qquad \qquad \qquad  \;
-q^{ \bigl((r-1)j+(b+c-1)/2\bigr)\bigl(rj+a\bigr)}
\Mults{L}{\frac{1}{2}(L-a-b)-rj}\right\}.
\label{the2}
\end{eqnarray}
\end{theorem}

To proof this, we note
that (\ref{the2}) satisfies (\ref{rr1}), thanks to
\begin{equation}
\Mult{N}{m}
= \Mult{N-1}{m-1}+q^{m}\Mult{N-1}{m}.
\label{qexp1}
\end{equation}
Similarly,
the proof that (\ref{the2}) satisfies (\ref{rr2}) follows by
application of
\begin{equation}
\Mult{N}{m}=\Mult{N-1}{m}+q^{N-m}\Mult{N-1}{m-1} .
\label{qexp2}
\end{equation}
To show that the initial condition (\ref{init}) holds,
note (\ref{qpoly}) as well as the range of $a$ and $b$.
This gives $j=0$ as the only non-vanishing
term in the sum, and hence $X_0(a,b,c) = \delta_{a,b}$.
Finally, $X_L(a,0)=0$ follows from (\ref{the2}) upon
substitution of $b=0$, $c=1$ and making the change
of variables $j\to -j$ in the first term
withing the curly braces. Analogously, $Y_L(a,r)$ follows from
(\ref{the2}) upon substituting $b=r$, $c=r-1$
and making the change of variables $j\to -j-1$ in the first term within
the braces. $\Box$.

To obtain the desired expression for $X_L(a,b)$, we set
$c=b+1$ in (\ref{the2}) yielding
\begin{eqnarray}
\lefteqn{
X_L(a,b) = f_{a,b}^{-1} \sum_{j=-\infty}^{\infty} \left\{
q^{j\bigl(r(r-1)j+rb-(r-1)a\bigr)}
\Mults{L}{\frac{1}{2}(L+a-b)-rj} \right. }
\nonumber \\
& & \left. \qquad \qquad \qquad \qquad \;
-q^{ \bigl((r-1)j+b\bigr) \bigl(rj+a\bigr)}
\Mults{L}{\frac{1}{2}(L-a-b)-rj} \right\}.
\label{cor2}
\end{eqnarray}
Combined with theorem~\ref{pageth1} on page~\pageref{pageth1},
this proves
Melzer's polynomial identities (\ref{Mid}).

In conclusion to this section we make two remarks about the
solution (\ref{cor2}).
First, the introduction of the auxiliary function $Y_L(a,b)$
could have been avoided, since from the definition (\ref{confsums})
one can obtain recurrences that only
involve the function $X_L(a,b)$. In particular,
\begin{equation}
X_L(a,b) =\left\{
\begin{array}{ll}
q^{L/2} X_{L-1}(a,b-1) + q^{(L-1)/2} X_{L-1}(a,b+1) & \\
\qquad \qquad \qquad \qquad \, + (1-q^{L-1}) X_{L-2}(a,b) \quad &
b=1,\ldots,r-3 \label{rr3} \\
q^{L/2} X_{L-1}(a,b-1) + X_{L-2}(a,b)  &  b=r-2,
\end{array} \right.
\end{equation}
with the same conditions on $X_L(a,b)$ as in (\ref{init})
and (\ref{bound}).
The price to be paid for this is that, in order
to show (\ref{cor2})
solves these relations, we need double application of (\ref{qexp1})
and (\ref{qexp2}). Interestingly though, in terms of the fermionic
left-hand side of (\ref{Mid}), the recurrences (\ref{rr3})
seem to be more natural, see e.g., ref.~\cite{Berkovich}.

A second remark we wish to make is that like the fermionic
result (\ref{Mid}), also (\ref{cor2}) has a nice interpretation
in terms of restricted lattice paths.
To see this, note that in order to obtain the generating
function for the restricted lattice paths,
we can first compute the generating function
$G_L(\emptyset)$
of restricted lattice paths without the restriction
$0\leq y\leq r-2$.
Since all paths that
go below $y=0$ and above $y=r-2$ have now incorrectly been
included, we have to
subtract the generating function $G_L(\downarrow)$ of paths
that somewhere go below $y=0$, as well as the generating function
$G_L(\uparrow)$ of paths that somewhere go above $y=r-2$.
However, we are again in error, since paths that go below $y=0$
as well as above $y=r-2$ have been subtracted twice.
To correct this we add $G_L(\downarrow,\uparrow)$
and $G_L(\uparrow,\downarrow)$, being the generating function
of all paths that somewhere go above $y=r-2$ {\em after} having
gone below $y=0$ and  the generating function of all paths
that somewhere go below $y=0$ {\em after} having gone
above $y=r-2$. Again this is no good, and we keep continuing
the process of adding and subtracting generating functions.
In formula this reads
\begin{equation}
X_L(a,b) = \sum_{j\geq 0} \Bigl\{
G_L(\underbrace{\downarrow\uparrow\cdots\downarrow\uparrow}_{2j})
+
G_L(\underbrace{\uparrow\downarrow\cdots\uparrow\downarrow}_{2j+2})
-
G_L(\underbrace{\downarrow\uparrow\cdots\uparrow\downarrow}_{2j+1})
-
G_L(\underbrace{\uparrow\downarrow\cdots\downarrow\uparrow}_{2j+1})
\Bigr\},
\label{sumG}
\end{equation}
with
$G_L(\underbrace{\downarrow\uparrow\cdots\downarrow\uparrow}_{2j})$
the generating function of all lattice paths that contain a sequence
of extrema
$\{(x_1,y_1),(x_2,y_2),\ldots,(x_{2j},y_{2j})\}$, with
$x_j>x_k$ for $j>k$,
$y_{2k-1}<0$ and $y_{2k}>r-2$, and with the other generating functions
defined similarly.
Of course, since we consider paths of fixed, finite length,
the above series only contains a finite number of nonzero terms.
Computing the functions $G_L$, we obtain
\begin{eqnarray}
G_L(\underbrace{\downarrow\uparrow\cdots\downarrow\uparrow}_{2j})
&=& f_{a,b}^{-1} \: q^{j\bigl(r(r-1)j-rb+(r-1)a\bigr)}
\Mults{L}{\frac{1}{2}(L-a+b)-rj} \nonumber \\
G_L(\underbrace{\uparrow\downarrow\cdots\uparrow\downarrow}_{2j})
&=& f_{a,b}^{-1} \: q^{j\bigl(r(r-1)j+rb-(r-1)a\bigr)}
\Mults{L}{\frac{1}{2}(L+a-b)-rj} \nonumber \\
G_L(\underbrace{\downarrow\uparrow\cdots\uparrow\downarrow}_{2j+1})
&=& f_{a,b}^{-1} \: q^{ \bigl((r-1)j+b\bigr) \bigl(rj+a\bigr)}
\Mults{L}{\frac{1}{2}(L-a-b)-rj} \label{compG} \\
G_L(\underbrace{\uparrow\downarrow\cdots\downarrow\uparrow}_{2j+1})
&=& f_{a,b}^{-1} \: q^{ \bigl((r-1)(j+1)-b\bigr) \bigl(r(j+1)+a\bigr)}
\Mults{L}{\frac{1}{2}(L+a+b)-r(j+1)} ,
\end{eqnarray}
for all $j\geq 0$.
Substitution in (\ref{sumG}) correctly reproduces the expression
(\ref{cor2}).
We remark that the above described method for computing
$X_L(a,b)$ is merely a rewording of the {\em sieving} technique
developed by Andrews in the context of partition theory,
see e.g., ref.~\cite{Andrews}. For the details of the calculation
leading to (\ref{compG}) we refer the reader to ref.~\cite{Andrews},
Chapter~9, and ref.~\cite{ABBBFV}.

\nsection{Summary and discussion}
In this paper we have, using the combinatorial
technique developed in part I, computed all
one-dimensional configuration sums of the $(r-1)$-state
ABF model. In contrast to the earlier results of
Andrews, Baxter and Forrester, our expressions are
of so-called fermionic type, and provide a new
proof of polynomial identities conjectured by Melzer.
In the limit of an infinitely large lattice, these
identities imply the fermionic
expressions for the $\chi_{b,a}^{(r-1,r)}(q)$
Virasoro characters as conjectured by the
Stony Brook group.
Using the Andrews--Bailey construction, we also
proved fermionic expressions
for several non-unitary minimal Virasoro characters.

In conclusion to this paper we make a few comments.
First, motivated by the ground breaking papers of the Stony Brook
group \cite{KKMMa,KKMMb,DKKMM,KM,DKMM},
a vast literature has arisen containing numerous
claims for identities of the
Rogers--Ramanujan type~[10,27-34]. We expect that
our fermionic method for computing generating
functions of restricted lattice paths can be applied to obtain
proof of several of these conjectures.
Other recently developed approaches towards
either proof, or an increase of understanding, of
Fermi-Bose character identities,
can for e.g., be found in refs.~[6,7,11,35-47].

A second remark is that in the $q=1$ limit, Melzer's
identities reduce to identities for the
number of $L$-step paths on the A$_{r-1}$ Dynkin
diagram, with fixed initial and final position.
Viewed this way, it turns out that Melzer's identities
are in fact a special 1-dimensional case of polynomial
identities for $q$-deformed path-counting on arbitrary
$d$-dimensional cuboids. In the limit of infinitely
long paths, these ``cuboid'' identities decouple
into products of Virasoro character identities.
The simplest example beyond Melzer's case is the
$q$-deformation of a path-counting formula on
a ``railroad'' of length $r-3$, reading:
\begin{eqnarray}
\lefteqn{
f_{a,b} \sum_{\vec{m} \equiv \vec{Q}_{a,b}}
q^{\: \case{1}{4} \,  \vec{m}^T C \: \vec{m}
-\case{1}{2} \, m_{r-a-1} }
\Mult{L}{m_1}
\prod_{j=2}^{r-3}
\Mult{\frac{1}{2}({\cal I} \, \vec{m}  + \vec{\e}_{r-a-1}
+\vec{\e}_{r-b-1})_j}{m_j} } \nonumber \\
& & =
\sum_{j=-\infty}^{\infty} \left\{
q^{j\big(r(r-1)j+rb-(r-1)a\big)}
\Mults{L}{2(r-1)j+b-a}_2
-q^{\big((r-1)j+b\big)\big(rj+a\big)}
\Mults{L}{2(r-1)j+b+a}_2
\right\},
\end{eqnarray}
for $a,b=1,\ldots,r-2$ and $L\geq 0$.
Here $\Mults{N}{m}_2$ are $q$-deformed trinomial coefficients
defined as \cite{AB}
\begin{equation}
\renewcommand{\arraystretch}{1.5}
\Mult{N}{m}_2 =
\sum_{k\geq 0} q^{k(k+m)}
\Mult{N}{k} \Mult{N-k}{k+m}.
\end{equation}
A discussion for the case of arbitrary cuboids
will be presented elsewhere~\cite{cuboid}.

A final remark is that the result (\ref{Xferm}) proven in this paper
has nice partition theoretical interpretations.
One follows from the work of Andrews {\em et al.}~\cite{ABBBFV},
stating that the one-dimensional configuration sum $X_L(a,b)$
is the generating function of all partitions into at most
$\case{1}{2}(L+a-b)$ parts, each part $\leq \case{1}{2}(L-a+b)$,
such that the hook differences on the $(1-b)$-th diagonal are
$\geq b-a+1$ and on the $(r-b-2)$-th diagonal $\leq b-a$.
Another interpretation follows by viewing a restricted lattice path
with total energy $E$
as a partition of $2 E=\la_1+\la_2+\ldots + \la_M$, with
$\la_j$ the $j$-th $x$-position counted from the right,
where the path has no extremum.
With this map from paths to partitions, $X_L(a,b;q^2)$
is the generating function of partitions into parts
$\la_1,\la_2,\ldots,\la_M$, with $\la_M<\ldots <\la_2<\la_1\leq L$,
and
\begin{equation}
\begin{array}{l}
1-b \leq u_j -d_j \leq r-b-2 \qquad \quad \forall j=1,\ldots,M \\
u_M-d_M = a-b,\; a-b-1.
\end{array}
\end{equation}
Here $u_j$ is the number of parts $\la_k \equiv a-b+k$ for $k\leq j$
and
$d_j$ is the number of parts $\la_k \not\equiv a-b+k$ for $k\leq j$.

\section*{Acknowledgements}
I wish to thank Alexander~Berkovich, Omar~Foda,
Peter~Forrester and Barry~McCoy
for many interesting and helpful discussions.
This work is supported by the Australian Research Council.

\section*{Note added}
After completing this manuscript we received a
preprint by A.~Schilling~\cite{Schilling},
in which (\ref{Mid}) as well
as (\ref{Mid}) are proven as special cases of polynomial
identities for finitized branching functions
of the cosets

\appendix

\nsection{Proof of equivalence of (3.32) and (3.33)}
In this appendix we proof that (\ref{ds}) can be simplified to
yield the final result (\ref{ss}) for the one-dimensional
configuration sums.

The problem with this step is that the following statement
turns out to be false:
\begin{equation}
\sum_{\mu=0}^{\min(a,b)-1}
\left. \sum_{\vec{m}}
\right.^{(\mu)}
=\left.\sum_{\vec{m}}
\right.^{'},
\end{equation}
where $(\mu)$ denotes the restriction (\ref{rmmu}) and the prime the
restriction (\ref{rmprime}).
In particular, the number of vectors $\vec{m}$ which are in accordance
with the restriction (\ref{rmprime}) exceed the number over
vectors $\vec{m}$ in accordance with the restriction
(\ref{rmmu}) summed over $\mu$.
What we will show now is that each of the additional $\vec{m}$'s
allowed by the sum on the right-hand side gives a vanishing
contribution.

Let us assume that $\min(a,b)=M+1$, so that we have
$M+1$ terms in the sum over $\mu$. Let us further define $S_{\mu}$
as the set of $\vec{m}$-vectors allowed by the restriction (\ref{rmmu}).
In other words, $\vec{m}\in S_{\mu}$ if
\begin{equation}
\left\{
\begin{array}{l}
m_1,\ldots,m_{r-M-3} \geq M-\mu \\
m_{r-M+k-3} \geq M-\mu-k+1 \qquad k=1,\ldots , M-\mu \\
m_{r-\mu-2}= \ldots = m_{r-3}=0 \\
m_j\equiv \min(a-1,r-j-2)+\min(b-1,r-j-2)
\qquad j=1,\ldots, r-3.
\end{array} \right.
\label{Su}
\end{equation}
Note that $S_{\mu} \cap S_{\nu} = \emptyset$ for $\mu\neq \nu$.
We now use that $\Mults{N}{m}$ is non-vanishing for
$0\leq m \leq N$ only.
{}From this and the summand of (\ref{ds}), we infer
\begin{equation}
0\leq m_j \leq \frac{1}{2}\,\bigl(m_{j-1}+m_{j+1} +
\delta_{r-a-1,j} + \delta_{r-b-1,j}\bigr)
\qquad j=2,\ldots,r-3,
\label{mconds}
\end{equation}
with $m_{r-2}=0$. From this one immediately sees that
\begin{equation}
m_j \geq m_k \qquad \mbox{for} \quad j<k.
\label{order}
\end{equation}
However, interestingly enough,
this condition is not yet good enough for our purposes.
Instead we need to use the fact that $\min(a,b)=M+1$.
This combined with (\ref{mconds}) gives rise to
the following ordering for all $j=r-M-1,\ldots,r-3$:
\begin{equation}
\mbox{if } m_j\geq 1 \Rightarrow
m_{j-k} \geq k+1 \qquad \quad k=1,\ldots,M-r+j+2,
\label{so}
\end{equation}
in addition to (\ref{order}).

In the following we use the notation $A \cup B = C$
to mean $A \cup B\subseteq C$ and
\begin{equation}
\sum_{\vec{m} \in A} \sum_{\vec{m} \in B} f(\vec{m}) =
\sum_{\vec{m} \in C} f(\vec{m}),
\end{equation}
with $f(\vec{m})$ the summand in (\ref{ds}).

We now proceed by induction.
We set $T_n=T_{n-1} \cup S_{M-n}$,
with $T_0 = S_M$ and claim that $T_n$ is given by
\begin{equation}
\left\{ \begin{array}{l}
m_1,\ldots,m_{r-M+n-3} \geq 0 \\
m_{r-M+n-2}= \ldots = m_{r-3}=0 \\
m_j\equiv \min(a-1,r-j-2)+\min(b-1,r-j-2).
\end{array} \right.
\label{Tn}
\end{equation}
For $n=0$ this is correct by construction.
To show the induction step set
$\mu=M-n-1$ in (\ref{Su}).
This yields
\begin{equation}
\left\{ \begin{array}{l}
m_1,\ldots,m_{r-M-3} \geq n+1 \\
m_{r-M+k-3} \geq n-k+2 \qquad k=1,\ldots , n+1 \\
m_{r-M+n-1}= \ldots = m_{r-3}=0 \\
m_j\equiv \min(a-1,r-j-2)+\min(b-1,r-j-2).
\end{array} \right.
\end{equation}
Using the conditions (\ref{order}) and (\ref{so}) with $j=r-M+n-2$,
we can combine the above two equations to find that $T_{n+1}$
is given by (\ref{Tn}) with $n$ replaced by $n+1$.
This proves our claim (\ref{Tn}) and we obtain the
desired expression for $T_M$ by setting $n=M$ in (\ref{Tn}).
This indeed gives the restriction (\ref{rmprime}) we set out
to prove. $\Box$

\end{document}